%% file: main.tex
\documentclass[ALICE,manyauthors]{cernphprep}
\usepackage{newtxtext}
\usepackage[slantedGreek]{newtxmath}
\usepackage[comma,square,numbers,sort&compress]{natbib}
\usepackage{hyperref}
\usepackage{lineno}
\usepackage{xspace}
\usepackage{comment}
\usepackage{multirow}
\usepackage{hyphenat}
\usepackage{bm}
\usepackage[T1]{fontenc} 
\usepackage{orcidlink}

\begin{document}
\input{commands.tex}

\begin{titlepage}
\PHyear{2023}       
\PHnumber{005}      
\PHdate{23 January}  

\title{Measurement of the fraction of jet longitudinal momentum carried by \Lc baryons in pp collisions}
\ShortTitle{In-jet \Lc production in pp collisions at $\sqrt{s} = 13~\TeV$} 

\Collaboration{ALICE Collaboration\thanks{See Appendix~\ref{app:collab} for the list of collaboration members}}
\ShortAuthor{ALICE Collaboration} 

\begin{abstract}
\input{0_Abstract}
\end{abstract}
\end{titlepage}

\setcounter{page}{2} 

\input{1_Introduction}

\input{2_Apparatus_Data}

\input{3_Analysis}

\input{3b_Systematics}

\input{4_Results}

\input{5_Conclusion}

\newenvironment{acknowledgement}{\relax}{\relax}
\begin{acknowledgement}
\section*{Acknowledgements}
\input{fa_2022-11-21_Opt_C.tex}
\end{acknowledgement}

\bibliographystyle{utphys}   
\bibliography{bibliography}

\newpage
\appendix

%
%

\section{The ALICE Collaboration}
\label{app:collab}
\input{2022-11-21-Alice_Authorlist_2022-11-21_Opt_C.tex}
\end{document}

%% file: commands.tex
%

\newcommand{\pp}           {pp\xspace}
\newcommand{\pplong}       {proton-proton\xspace}
\newcommand{\ppbar}        {\mbox{$\mathrm {p\overline{p}}$}\xspace}
\newcommand{\ee}           {\mbox{$\mathrm {e^{+}e^{-}}$}\xspace}
\newcommand{\ep}           {\mbox{$\mathrm {e^{-}p}$}\xspace}
\newcommand{\XeXe}         {\mbox{Xe--Xe}\xspace}
\newcommand{\PbPb}         {\mbox{Pb--Pb}\xspace}
\newcommand{\pA}           {\mbox{pA}\xspace}
\newcommand{\pPb}          {\mbox{p--Pb}\xspace}
\newcommand{\AuAu}         {\mbox{Au--Au}\xspace}
\newcommand{\dAu}          {\mbox{d--Au}\xspace}
\let\eepm=\ee
\newcommand{\hi}           {heavy-ion\xspace}
\newcommand{\Hi}           {Heavy-ion\xspace}

\newcommand{\s}            {\ensuremath{\sqrt{s}}\xspace}
\newcommand{\snn}          {\ensuremath{\sqrt{s_{\mathrm{NN}}}}\xspace}
\newcommand{\pt}           {\ensuremath{p_{\rm T}}\xspace}
\newcommand{\ptjet}        {\ensuremath{p_{\rm T}^{\rm jet\;ch}}\xspace}
\newcommand{\kt}           {\ensuremath{k_{\rm T}}\xspace}
\newcommand{\meanpt}       {$\langle p_{\mathrm{T}}\rangle$\xspace}
\newcommand{\ycms}         {\ensuremath{y_{\rm CMS}}\xspace}
\newcommand{\ylab}         {\ensuremath{y_{\rm lab}}\xspace}
\newcommand{\etarange}[1]  {\mbox{$\left | \eta \right |~<~#1$}\xspace}
\newcommand{\yrange}[1]    {\mbox{$\left | y \right |~<~#1$}\xspace}
\newcommand{\dndy}         {\ensuremath{\mathrm{d}N_\mathrm{ch}/\mathrm{d}y}\xspace}
\newcommand{\dndeta}       {\ensuremath{\mathrm{d}N_\mathrm{ch}/\mathrm{d}\eta}\xspace}
\newcommand{\avdndeta}     {\ensuremath{\langle\dndeta\rangle}\xspace}
\newcommand{\dNdy}         {\ensuremath{\mathrm{d}N_\mathrm{ch}/\mathrm{d}y}\xspace}
\newcommand{\Npart}        {\ensuremath{N_\mathrm{part}}\xspace}
\newcommand{\Ncoll}        {\ensuremath{N_\mathrm{coll}}\xspace}
\newcommand{\dEdx}         {\ensuremath{\textrm{d}E/\textrm{d}x}\xspace}
\newcommand{\RpPb}         {\ensuremath{R_{\rm pPb}}\xspace}
\newcommand{\DeltaM}       {\ensuremath{\Delta M}\xspace}
\let\dNchdeta=\avdndeta
\newcommand{\ntrkl}        {\ensuremath{N_\mathrm{trkl}}\xspace}
\newcommand{\vzeromperc}   {\ensuremath{p_\mathrm{V0M}}\xspace}
\newcommand{\zvtx}         {\ensuremath{z_\mathrm{vtx}}\xspace}
\newcommand{\raa}          {\ensuremath{R_\mathrm{AA}}\xspace}
\let\RAA=\raa
\newcommand{\z}          {\ensuremath{z^{\mathrm{ch}}_\mathrm{||}}\xspace}

\newcommand{\sqrts}        {\ensuremath{\sqrt{s}}\xspace}
\newcommand{\sqrtsNN}      {\ensuremath{\sqrt{s_{\mathrm{NN}}}}\xspace}
\newcommand{\nineH}        {$\sqrt{s}=0.9$~Te\kern-.1emV\xspace}
\newcommand{\seven}        {$\sqrt{s}=7$~Te\kern-.1emV\xspace}
\newcommand{\twoH}         {$\sqrt{s}=0.2$~Te\kern-.1emV\xspace}
\newcommand{\twosevensix}  {$\sqrt{s}=2.76$~Te\kern-.1emV\xspace}
\newcommand{\five}         {$\sqrt{s}=5.02$~Te\kern-.1emV\xspace}
\newcommand{\thirteen}     {$\sqrt{s}=13$~Te\kern-.1emV\xspace}
\newcommand{\twosevensixnn}{$\sqrt{s_{\mathrm{NN}}}=2.76$~Te\kern-.1emV\xspace}
\newcommand{\fivenn}       {$\sqrt{s_{\mathrm{NN}}}=5.02$~Te\kern-.1emV\xspace}
\newcommand{\LT}           {L{\'e}vy-Tsallis\xspace}
\newcommand{\GeVc}         {\ensuremath{\mathrm{GeV}/c}\xspace}
\newcommand{\MeVc}         {\ensuremath{\mathrm{MeV}/c}\xspace}
\newcommand{\TeV}          {\ensuremath{\mathrm{TeV}}\xspace}
\newcommand{\GeV}          {\ensuremath{\mathrm{GeV}}\xspace}
\newcommand{\MeV}          {\ensuremath{\mathrm{MeV}}\xspace}
\newcommand{\GeVcc}        {\ensuremath{\mathrm{GeV}/c^{2}}\xspace}
\newcommand{\MeVcc}        {\ensuremath{\mathrm{MeV}/c^{2}}\xspace}
\newcommand{\lumi}         {\ensuremath{\mathcal{L}}\xspace}
\newcommand{\degree}       {\ensuremath{^{\rm o}}\xspace}
\newcommand{\inversemub}   {\ensuremath{\mu \rm b^{\rm -1}}\xspace}

\let\gevc=\GeVc
\let\mevc=\MeVc
\let\tev=\TeV
\let\gev=\GeV
\let\mev=\MeV
\let\gevcc=\GeVcc
\let\mevcc=\MeVcc
\let\GeVmom=\GeVc
\let\MeVmom=\MeVc
\let\GeVmass=\GeVcc
\let\MeVmass=\MeVcc

\newcommand{\ITS}          {\rm{ITS}\xspace}
\newcommand{\TOF}          {\rm{TOF}\xspace}
\newcommand{\ZDC}          {\rm{ZDC}\xspace}
\newcommand{\ZDCs}         {\rm{ZDCs}\xspace}
\newcommand{\ZNA}          {\rm{ZNA}\xspace}
\newcommand{\ZNC}          {\rm{ZNC}\xspace}
\newcommand{\SPD}          {\rm{SPD}\xspace}
\newcommand{\SDD}          {\rm{SDD}\xspace}
\newcommand{\SSD}          {\rm{SSD}\xspace}
\newcommand{\TPC}          {\rm{TPC}\xspace}
\newcommand{\TRD}          {\rm{TRD}\xspace}
\newcommand{\VZERO}        {\rm{V0}\xspace}
\newcommand{\VZEROA}       {\rm{V0A}\xspace}
\newcommand{\VZEROC}       {\rm{V0C}\xspace}

\newcommand{\pion}{\ensuremath{\uppi}}
\newcommand{\pip}          {\ensuremath{\pion^{+}}\xspace}
\newcommand{\pim}          {\ensuremath{\pion^{-}}\xspace}
\newcommand{\kap}          {\ensuremath{\rm{K}^{+}}\xspace}
\newcommand{\kam}          {\ensuremath{\rm{K}^{-}}\xspace}
\newcommand{\pbar}         {\ensuremath{\rm\overline{p}}\xspace}
\newcommand{\kzero}        {\ensuremath{\textnormal{K}^{0}_{\textnormal{S}}}\xspace}
\newcommand{\lmb}          {\ensuremath{\upLambda}\xspace}
\newcommand{\almb}         {\ensuremath{\overline{\upLambda}}\xspace}
\newcommand{\Om}           {\ensuremath{\Omega^-}\xspace}
\newcommand{\Mo}           {\ensuremath{\overline{\Omega}^+}\xspace}
\newcommand{\X}            {\ensuremath{\Xi^-}\xspace}
\newcommand{\Ix}           {\ensuremath{\overline{\Xi}^+}\xspace}
\newcommand{\Xis}          {\ensuremath{\Xi^{\pm}}\xspace}
\newcommand{\Oms}          {\ensuremath{\Omega^{\pm}}\xspace}
\newcommand{\Vdecay} 	   {\ensuremath{\rm V^{0}}\xspace}
\newcommand{\Kzeros}       {\ensuremath{\textnormal{K}^{0}_{\textnormal{S}}}\xspace}

\newcommand{\Dzero}        {\ensuremath{\rm D^{0}}\xspace}
\newcommand{\Dplus}        {\ensuremath{\rm D^{+}}\xspace}
\newcommand{\Dstar}        {\ensuremath{\rm D^{*+}}\xspace}
\newcommand{\Ds}           {\ensuremath{\rm D^{+}_{\rm s}}\xspace}
\newcommand{\Lambdac}      {\ensuremath{\upLambda_{\textnormal{c}}^{+}}\xspace}
\newcommand{\Lambdab}      {\ensuremath{\upLambda_{\textnormal{b}}^{0}}\xspace}
\let\Lc=\Lambdac
\let\Lb=\Lambdab

\newcommand{\DtoKpi}          {\ensuremath{\rm D^{0} \to K^{-}\pi^{+}}\xspace}
\newcommand{\Dstophip}        {\ensuremath{\rm D_{\rm s}^{+} \to \phi \pi^{+} \to K^{+}K^{-}\pi^{+}}\xspace}
\newcommand{\LctopKpi}        {\ensuremath{\upLambda_{\textnormal{c}}^{+} \to \textnormal{p}\textnormal{K}^{-}\pion^{+}}\xspace}
\newcommand{\LctopKzeros}     {\ensuremath{\upLambda_{\textnormal{c}}^{+} \to \textnormal{p} \Kzeros}\xspace}
\newcommand{\Lcpm}       {\ensuremath{\upLambda_{\textnormal{c}}^{\pm}}\xspace}
\newcommand{\LctopKzerosfull} {\ensuremath{\upLambda_{\textnormal{c}}^{+} \to \textnormal{p} \Kzeros \to \textnormal{p}\pion^{+}\pion^{-}}\xspace}
\newcommand{\pKpi}            {\ensuremath{\rm pK^{-}\pi^{+}}\xspace}
\newcommand{\pKzeros}         {\ensuremath{\rm pK^{0}_{\rm S}}\xspace}
\let\LambdactopKpi=\LctopKpi
\let\LambdactopKzeros=\LctopKzeros
\let\LambdactopKzerosfull=\LctopKzerosfull

\newcommand{\ptLc}         {\ensuremath{p_{\textnormal{T}}^{\Lc}}\xspace}

\newcommand{\DsDzero}         {\ensuremath{\Ds/\Dzero}\xspace}
\newcommand{\LcDzero}         {\ensuremath{\Lc/\Dzero}\xspace}
\let\LcD=\LcDzero
\newcommand{\inel}            {\ensuremath{\rm INEL_{>0}}\xspace}
\let\INEL=\inel

\newcommand{\Hf}            {Heavy-flavour\xspace}
\newcommand{\hf}            {heavy-flavour\xspace}

\newcommand{\alphas}        {\ensuremath{\alpha_\mathrm{S}\xspace}}

\newcommand{\secletter}[1]    {\vspace{0.3cm}\textit{#1}}

\newcommand{\RAAprompt}    {\ensuremath{R^{\rm prompt~\Lc}_\mathrm{AA}}\xspace}
\newcommand{\RAAnonprompt}    {\ensuremath{R^{\rm non-prompt~\Lc}_\mathrm{AA}}\xspace}


\newcommand{\Rres}          {\rm{R}\xspace}

%% file: 0_Abstract.tex
Recent measurements of charm-baryon production in hadronic collisions have questioned the universality of charm-quark fragmentation across different collision systems.
In this work the fragmentation of charm quarks into charm baryons is probed, by presenting the first measurement of the longitudinal jet momentum fraction carried by \Lambdac baryons, \z, in hadronic collisions. The results are obtained in proton--proton (pp) collisions at $\s=13$ \TeV at the LHC, with \Lambdac baryons and charged (track-based) jets reconstructed in the transverse momentum intervals of $3 \leq \ptLc < 15$ \GeVc and $7 \leq \ptjet < 15$ \GeVc, respectively.
The \z distribution is compared to a measurement of \Dzero-tagged charged jets in \pp collisions as well as to PYTHIA 8 simulations. The data hints that the fragmentation of charm quarks into charm baryons is softer with respect to charm mesons, in the measured kinematic interval, as predicted by hadronisation models which include colour correlations beyond leading-colour in the string formation.

%% file: 1_Introduction.tex
Heavy-flavour hadrons are produced in high-energy particle collisions through the fragmentation of heavy (charm and beauty) quarks, which typically originate in hard scattering processes in the early stages of the collisions. The most common theoretical approach to describe heavy-flavour production in hadronic collisions is based on the quantum chromodynamics (QCD) factorisation approach~\cite{Collins:1989gx}, and consists of a convolution of three independent terms: the parton distribution functions of the incoming hadrons, the cross sections of the partonic scattering producing the heavy quarks, and the fragmentation functions that parametrise the evolution of a heavy quark into given species of heavy-flavour hadrons. As the transition of quarks to hadrons cannot be described in perturbation theory, the fragmentation functions cannot be calculated and must be extracted from data.

Fragmentation functions of charm quarks to charm baryons and mesons have been constrained in \ee~\cite{belle_charm_fragmentation,belle_charm_crosssection,aleph_charm_production} and \ep~\cite{zeus_charm_fragmentation,H1:2008nbh} collisions, using a variety of different observables, such as the hadron momentum as a fraction of its maximum possible momentum, as dictated by the centre-of-mass energy of the collision. Another method to probe the fragmentation of quarks to hadrons is to parametrise the hadron momentum in relation to the momentum of jets, which are collimated bunches of hadrons giving experimental access to the properties of the scattered quark.
Recently, the production of charm mesons in jets, probed via the fractional longitudinal momentum of the jet carried by the D meson, was measured in pp collisions at the Large Hadron Collider (LHC)~\cite{D0fragmentation,D0fragmentation7Tev,ATLAS:2011chi} and appears consistent with Monte Carlo (MC) simulations tuned on \ee data. These measurements support the assumption of fragmentation universality across collision systems in the charm-meson sector. This assumption underpins theoretical calculations describing the production of heavy-flavour hadrons in hadronic collisions, which make use of fragmentation functions tuned on \ee and \ep data.

Measurements of the production cross sections of baryons in \pp collisions have questioned the hypothesis of fragmentation universality across collision systems~\cite{ALICE:2016fzo}. In the charm sector, which provides a clean probe of hadronisation phenomena due to the large  mass of the charm quark, recent measurements performed by the ALICE Collaboration~\cite{Acharya:2017kfy, Acharya:2020uqi, LcD013TeV, Acharya:2020lrg,ALICE:2021npz,ALICE:2021dhb,ALICE:OmegaC,ALICE:XiC, ALICE:XiC2} in pp collisions have shown that the ratio of the \Lc (and other charm baryons) and \Dzero production cross sections measured at low \pt ($\lesssim 12$ \GeVc) is significantly larger than the value expected from MC simulations in which the charm fragmentation is tuned on \ee and \ep measurements, such as PYTHIA~8~\cite{Skands:2014pea} with the Monash tune~\cite{Sjostrand:2014zea} or HERWIG~7~\cite{Bahr:2008pv}. A recent measurement of the \LcD ratio in pp collisions, performed by the ALICE Collaboration in intervals of charged-particle multiplicity, also points to a substantial increase of the \LcD ratio with increasing multiplicity, with respect to \ee collisions, starting at very low multiplicities~\cite{ALICE:2021npz}.

The study of charm-baryon production in jets can provide more differential insights into hadronisation mechanisms in pp collisions, compared to $\pt$-differential cross sections and yield ratios of heavy-flavour hadrons, allowing for a more accurate study of the dynamical properties of baryon production. In this paper, the first measurement of the longitudinal momentum fraction of the jet carried by \Lambdac baryons, \z, is presented. The measurement is performed in \pp collisions at $\s=13$~\TeV in the interval $0.4 \leq \z \leq 1.0$. The \z distribution, fully corrected to particle level, is presented for prompt (charm-quark initiated) \Lambdac-tagged jets with $7 \leq \ptjet < 15$~\GeVc and $3 \leq \ptLc < 15$~\GeVc. The results are then compared to PYTHIA~8 simulations~\cite{Skands:2014pea,Christiansen:2015yqa}, including a version where mechanisms beyond the leading-colour approximation are considered in
string formation processes during hadronisation
~\cite{Sjostrand:2014zea}, and to an analogous measurement of the \z distribution  of \Dzero mesons, performed by the ALICE Collaboration~\cite{D0fragmentation}.

%% file: 2_Apparatus_Data.tex
A full description of the ALICE setup and apparatus can be found in Refs.~\citenum{Aamodt:2008zz,Abelev:2014ffa}. The main detectors used in this analysis are the Inner Tracking System (ITS), which is used for vertex reconstruction and tracking; the Time Projection Chamber (TPC), which is used for tracking and particle identification (PID); and the Time-Of-Flight (TOF) detector, which is used for PID. These detectors cover a pseudorapidity interval of $|\eta| < 0.9$. The analysis was performed on pp collisions at $\sqrt{s} = 13~\mathrm{TeV}$, collected using a minimum-bias (MB) trigger during the years 2016, 2017, and 2018. The trigger condition required coincident signals in the two scintillator arrays of the \VZERO detector, with background events originating from beam--gas interactions removed offline using timing information from the \VZERO. To mitigate against pile-up effects, events with multiple reconstructed primary vertices were rejected. To ensure uniform acceptance, only events with a primary-vertex position along the beam axis direction of $|z_{\rm vtx}|<10$ cm around the nominal interaction point were accepted. After the selections described above, the data sample consisted of $1.7\times10^{9}$ events, corresponding to an integrated luminosity of $\mathcal{L}_{\text{int}}=29~\mathrm{nb}^{-1}$~\cite{alice-luminosity}.

%% file: 3_Analysis.tex
The \Lambdac candidates and their charge conjugates were reconstructed via the hadronic $\LctopKzerosfull$ decay channel with a total branching ratio of $(1.10 \pm 0.06)\% $~\cite{Zyla:2020zbs}, in the \Lambdac transverse-momentum interval of $3 \leq \ptLc < 15$~\GeVc. Only tracks with $|\eta|<0.8$ and $\pt>0.4$~\gevc, which fulfilled the track quality selections described in Ref.~\citenum{Acharya:2020lrg}, were considered for the \Lambdac reconstruction. The \Lambdac candidates themselves were reconstructed in the $|y^{\Lambdac}|<0.8$ rapidity interval. The \Lambdac-candidate selection was performed using a multivariate technique based on the Boosted Decision Tree (BDT) algorithm provided by the XGBoost package~\cite{Chen:2016:XST:2939672.2939785}. The features considered in the optimisation include the PID signal for the proton track, the invariant mass of the \kzero-meson candidate, and topological variables that exploit the kinematic properties of the displaced \kzero-meson decay vertex.
The training was performed in intervals of \Lc-candidate \pt, considering prompt signal candidates from PYTHIA~8 events with the Monash tune~\cite{Skands:2014pea,Sjostrand:2014zea}, transported through a~realistic description of the detector geometry and material budget using GEANT~3~\cite{GEANT3}. Background candidates were extracted from the sidebands of the invariant-mass distributions in data. The probability thresholds of the BDT selections were tuned, using MC simulations, to maximise the statistical significance for the signal. Further details on the \Lambdac-candidate reconstruction and machine learning procedure are provided in Ref.~\citenum{ALICE:2021npz}, where the same reconstruction and BDT model were employed.

For the events where at least one selected \Lambdac candidate was identified, a jet-finding procedure was performed, using the FastJet package~\cite{Cacciari_2012}. Prior to jet clustering, the \Lambdac-candidate daughter tracks were replaced by the reconstructed \Lambdac-candidate four-momentum vector. Charged jet finding was carried out on charged tracks with $|\eta|<0.9$ and $\pt>0.15$~\gevc, using the anti-$k_{\rm T}$ algorithm~\cite{Cacciari_2008}, with a resolution parameter of $R=0.4$. Tracks were combined using the $E$-scheme recombination~\cite{EScheme}, with the jet transverse momentum limited to the interval of $5 \leq \ptjet <35$~\GeVc. The full jet cone was required to be within the ALICE central barrel acceptance, limiting the jet axis to the interval $|\eta_{\rm jet}| < 0.5$. Only jets tagged via the presence of a reconstructed \Lambdac candidate amongst their constituents were considered for the analysis. For events where more than one \Lambdac candidate was found, the jet finding and tagging pass was performed independently for each candidate, with only the daughters of that particular candidate replaced by the corresponding \Lambdac four-vector each time. In mechanisms of hadronisation that include colour correlations beyond the leading-colour approximation~\cite{Sjostrand:2014zea}, which have been shown to be relevant in hadronic collisions at LHC energies~\cite{ALICE:2016fzo}, hadrons can be formed in processes that combine quarks from the parton shower with those from the underlying event~\cite{Han:2016uhh}. As such, the underlying event is not well defined with respect to the measured hadron distributions. Therefore no underlying event correction is implemented in this work.

The fragmentation of charm quarks to \Lambdac baryons is probed by measuring the fraction of the jet momentum carried by the \Lambdac along the direction of the jet axis, \z. This is calculated for each jet using

\begin{equation}
    \z = \frac{\bm{p}_{\textnormal{jet}}\cdot\bm{p}_{\Lambdac}}{\bm{p}_{\textnormal{jet}}\cdot\bm{p}_{\textnormal{jet}}},
    \label{Eq:FF}
\end{equation}

where $\bm{p}_{\textnormal{jet}}$ and $\bm{p}_{\Lambdac}$ are the jet and \Lambdac three-momentum vectors, respectively.

The \z distributions of true \Lambdac-tagged charged jets were extracted in intervals of \Lambdac \pt and \ptjet using a~sideband subtraction procedure. To enact this subtraction, the invariant-mass ($m_{\rm inv}$) distributions of \Lambdac candidates, obtained for each \Lambdac \pt and \ptjet interval, were fitted with a~function comprising a~Gaussian for the signal and an exponential for the background. The fit parameters were then used to define signal (containing the majority of true signal candidates) and sideband (entirely composed of background candidates) regions, defined by $|m_{\rm inv} - \mu_{\rm fit}| < 2\sigma_{\rm fit}$ and $ 4\sigma_{\rm fit} < |m_{\rm inv} - \mu_{\rm fit}| < 9\sigma_{\rm fit}$, respectively, where $\mu_{\rm fit}$ and $\sigma_{\rm fit}$ represent the mean and sigma of the fitted Gaussian distributions. The \z(\ptLc,\ptjet) distributions were extracted in the signal and sideband regions, with the sideband distribution scaled by the ratio of the background function integrals in the signal and sideband regions. The sideband distribution was then subtracted from the signal one, with the resulting distribution scaled to account for the fact that the $2\sigma_{\rm fit}$ width of the signal region only encompasses approximately $95\%$ of the total signal, to obtain the sideband subtracted \z yield in each $\ptLc$ and \ptjet interval.

To account for the reconstruction and selection efficiency of the \Lambdac-tagged jet signal, the sideband subtracted \z distributions in each \ptLc and \ptjet interval, $N(\z,\ptLc,\ptjet)$,  were scaled by the reconstruction efficiency of prompt \Lambdac-tagged jets, $\epsilon_{\textnormal{prompt}}$, and summed over the entire \ptLc interval to obtain the efficiency-corrected \z yield of \Lambdac-tagged jets,  $N^{\textnormal{corr}}(\z,\ptjet)$, given by
\begin{equation}
    N^{\text{corr}}(\z, \ptjet) = \sum_{\ptLc} \frac{N(\z,\ptLc,\ptjet)}{\epsilon_{\rm{prompt}}(\ptLc)} .
    \label{Eq:pTDSum}
\end{equation}
The $\epsilon_{\textnormal{prompt}}(\ptLc)$ efficiency is strongly dependent on \ptLc, ranging from about $20\%$ at $3<\ptLc<4~\GeVc$ to $40\%$ at $12<\ptLc<24~\GeVc$, and was calculated using PYTHIA~8 simulations with the Monash tune containing prompt \Lambdac-tagged jets, transported through the detector using GEANT~3. This efficiency does not exhibit a \ptjet dependence.

In order to isolate the $N^{\text{corr}}(\z,\ptjet)$ distribution of prompt \Lambdac-tagged jets, a~feed-down subtraction was employed to remove the non-prompt (beauty-quark initiated) contribution. The non-prompt cross section was obtained from particle level POWHEG~\cite{POWHEG} + PYTHIA~6~\cite{Sjostrand:2006za} + EvtGen~\cite{evtgen} simulations, as a function of $\ptjet$, $\ptLc$ and $\z$, and was scaled according to the integrated luminosity of the analysed data sample and the branching ratio of the $\LctopKzerosfull$ decay channel. The resulting particle-level yield was multiplied by the ratio of the non-prompt to prompt \Lambdac-tagged jet reconstruction and selection efficiency in intervals of \ptLc and integrated over the \ptLc range. The simulated non-prompt results were then folded to reconstructed level, using a~four-dimensional response matrix generated using non-prompt \Lambdac-tagged jets in PYTHIA~8 with the Monash tune, transported through a simulation of the ALICE detector using GEANT~3. The response matrix was constructed as a function of \ptjet and \z at generator and reconstruction levels. The folded results were then subtracted from the measured $N^{\text{corr}}(\z, \ptjet)$ distribution in data, removing the non-prompt contribution. The estimated fraction of \Lambdac-tagged jets coming from b-quark fragmentation is found to be about 5$\%$, with no significant \z dependence.

A two-dimensional Bayesian unfolding procedure~\cite{unfolding} was performed to correct for detector effects and obtain the \z distribution for prompt \Lambdac-tagged jets at particle level. A four-dimensional response matrix as a function of \ptjet and \z, at generator and reconstruction levels, was populated with prompt \Lambdac-tagged jets, obtained with PYTHIA~8 simulations with the Monash tune, passed through a simulation of the ALICE detector using GEANT~3. The measured data and response matrix were provided in the intervals of $5 \leq \ptjet < 35$~\GeVc and $0.4 \leq \z \leq 1.0$, with the final unfolded results reported in the intervals $7 \leq \ptjet <15$ \GeVc and $0.4 \leq \z \leq 1.0$. The extended $\ptjet$ range includes two padding intervals for the unfolding from $5 \leq \ptjet < 7$~\GeVc and $15 \leq \ptjet < 35$~\GeVc, which allow the unfolding to account for migrations in and out of the reported $7 \leq \ptjet < 15$~\GeVc interval. Corrections accounting for migrating entries in and out of the response matrix ranges, as modelled by the same MC simulation, were also applied. The corrected \z distribution is normalised to the total number of \Lambdac-tagged jets in the reported \z and \ptjet interval.

%% file: 3b_Systematics.tex
The systematic uncertainties affecting the measurement were evaluated, in each \z interval, by modifying the strategy adopted at various steps of the analysis procedure and assessing the impact on the unfolded \z distribution. The total systematic uncertainty includes contributions from multiple sources. The first considered source is the sideband subtraction procedure, whose contribution (ranging from 3.7\% to 7.6\% depending on the \z inteval) was estimated by varying the invariant-mass fit parameters as well as the invariant-mass intervals of the signal and sideband regions. The contribution from the BDT selection of \Lambdac candidates (from 7.3\% to 19\%) was estimated by varying the BDT probability thresholds to induce a $25\%$ variation in the \Lambdac-tagged jet reconstruction and selection efficiency. The uncertainty from the jet energy resolution (from 4.5\% to 19\%) was estimated by recalculating the response matrix used for unfolding with a $4\%$ reduced tracking efficiency. The reduction in the tracking efficiency was evaluated by varying the track-selection criteria and propagating the ITS--TPC track-matching efficiency uncertainty.  
The uncertainty on the feed-down subtraction ($< 2\%$) was estimated by varying the choice of POWHEG parameters considered to generate the feed-down cross section, including the factorisation and renormalisation scales, as well as the mass of the beauty quark, which were varied according to theoretical prescriptions~\cite{powheg-variations}. Finally the contribution from the unfolding procedure (from 1.1\% to 2.7\%) was estimated by altering the choice of prior, regularisation parameter, and ranges of the response matrix. For each of the aforementioned categories, several variations were made and the root-mean-square of the resulting distributions was considered. The exceptions are related to the contribution associated to the choice of parameters of the POWHEG calculations, where only the largest deviation from the central result, in each direction, was considered, as well as the uncertainty on the jet energy resolution where the variation with respect to the central result was taken as the uncertainty.  All uncertainties (other than from the feed-down subtraction) were then symmetrised. The uncertainties were combined in quadrature to obtain the total systematic uncertainty on the measurement, which ranges from $13\%$ to $28\%$.

%% file: 4_Results.tex
\begin{figure}[tb!]
  \begin{center}
    \includegraphics[width=0.49\textwidth]{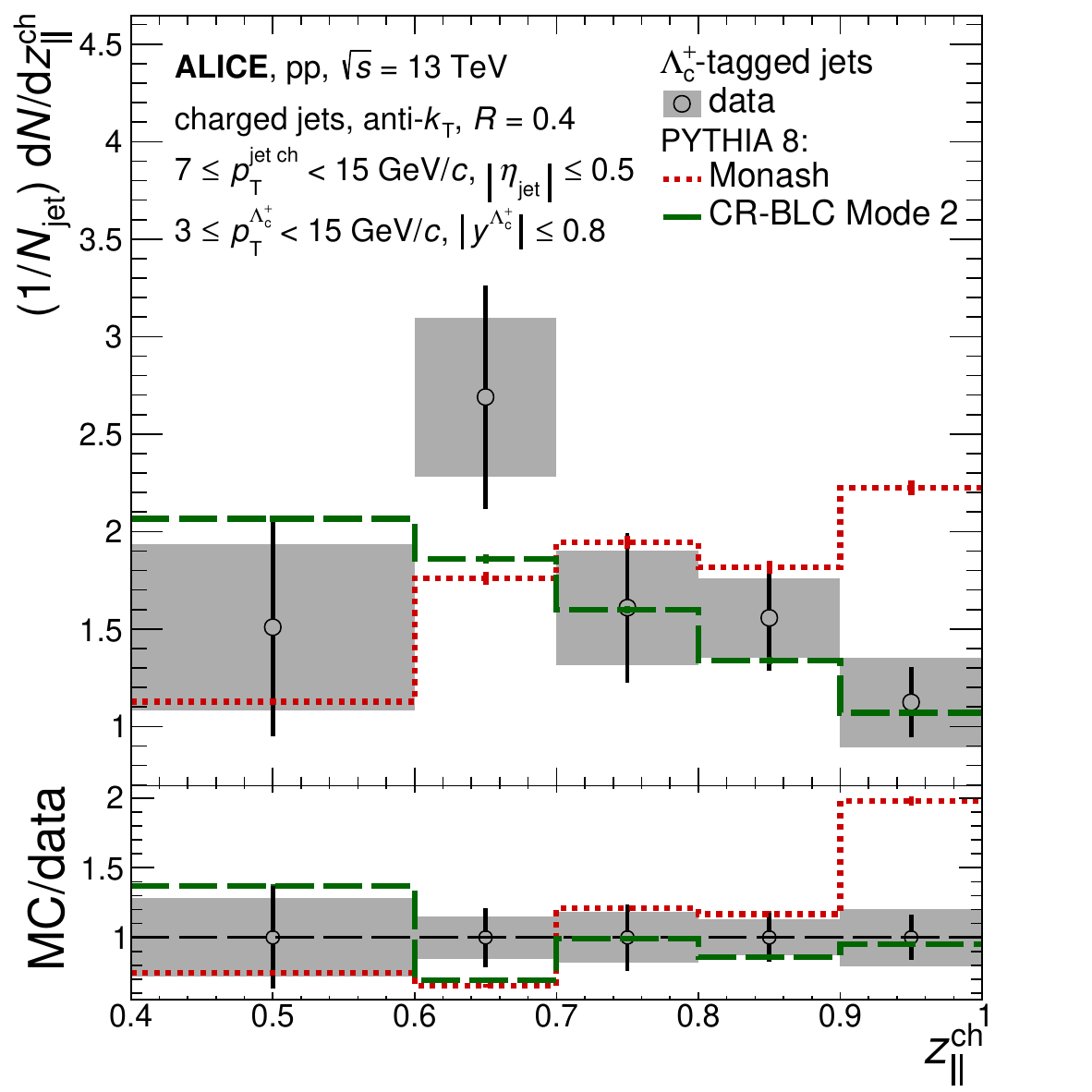}
    \includegraphics[width=0.49\textwidth]{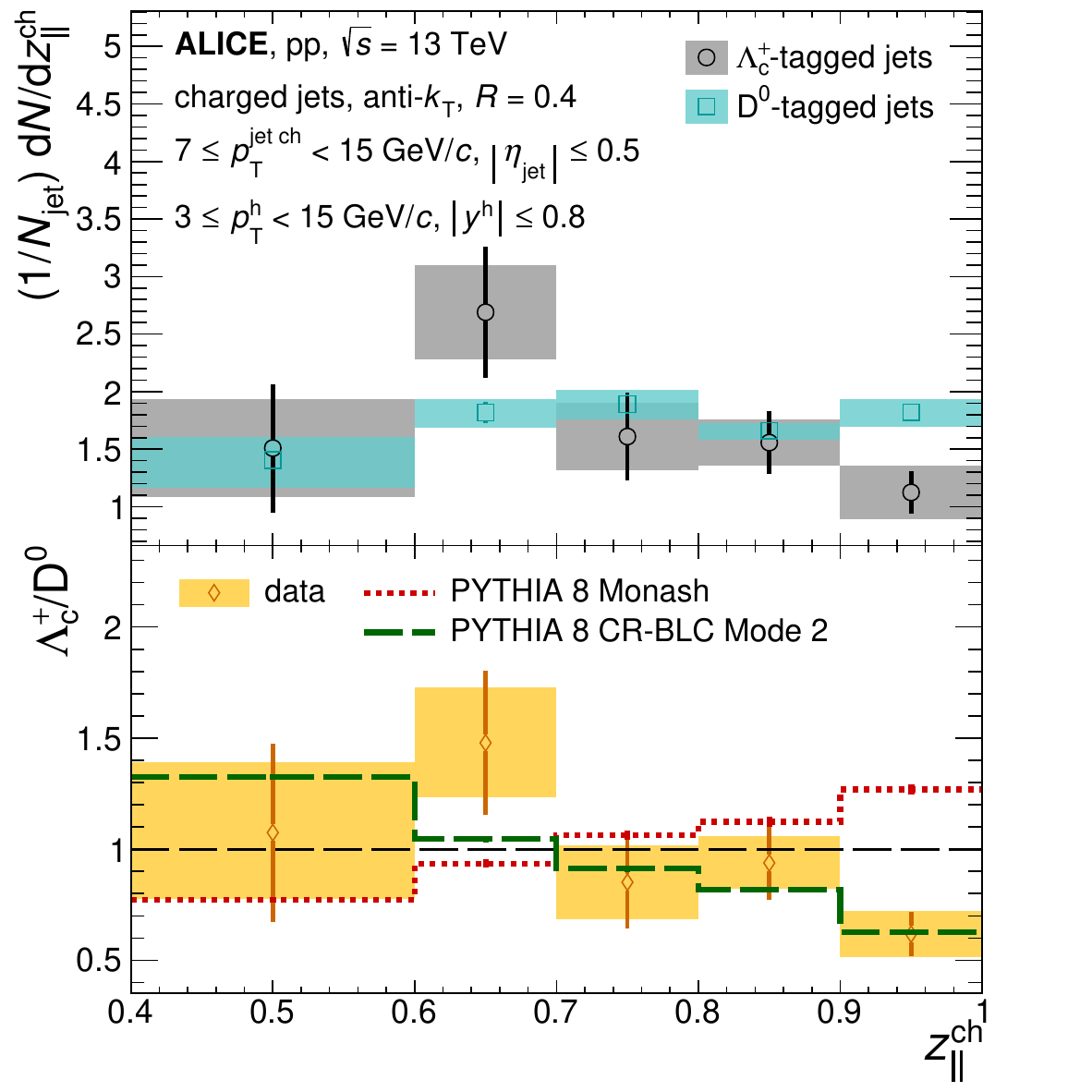}
    \caption{(Left) Fully corrected normalised \z distribution of \Lambdac-tagged charged jets (black open circles) measured in the $7\leq\ptjet<15$~\GeVc and $3\leq\ptLc<15$~\GeVc intervals in pp collisions at $\s = 13$~TeV, compared with predictions from different PYTHIA~8 tunes~\cite{Skands:2014pea,Sjostrand:2014zea,Christiansen:2015yqa} (red-dotted and green-dashed lines). The ratios of the MC simulations to the data are shown in the bottom panel. (Right) Comparison of the measured \z distribution of \Lambdac-tagged jets and the previously measured \z distribution of \Dzero-tagged jets~\cite{D0fragmentation}, obtained in the same kinematic interval. The ratio of the \z distribution of \Lambdac-tagged and \Dzero-tagged jets is shown in the bottom panel for both the data and the different PYTHIA tunes.}
    \label{fig:result}
  \end{center}
\end{figure}

The fully corrected \z distribution of prompt \Lambdac-tagged charged jets in the intervals of $7 \leq \ptjet < 15$~\GeVc and $3\leq\ptLc<15$~\GeVc is presented in the left-hand panel of Fig.~\ref{fig:result} and compared to PYTHIA~8 simulations with two different tunes. In PYTHIA~8 the Lund string model of fragmentation is employed, where endpoints are confined by linear potentials encoded in strings. For the case of heavy quarks, the Lund fragmentation function is modified to account for the slower propagation of the massive endpoints compared to their massless counterparts. The Monash tune (red-dotted line)~\cite{Skands:2014pea}, in which the charm fragmentation is tuned on \ee measurements, predicts a harder fragmentation than the measurement. An evaluation of the $\chi^{2}/$ndf between the measured data points and the model was performed, combining the statistical and systematic uncertainties on the data in quadrature and assuming the uncertainties are uncorrelated across the \z intervals. This exercise determines that there is a $0.4\%$ probability that the model describes the data. A better agreement is achieved by PYTHIA~8 with the CR-BLC Mode~2 tune, which includes colour reconnection mechanisms beyond the leading-colour approximation~\cite{Christiansen:2015yqa} (green-dashed line). In this model, the minimisation of the string potential is implemented considering the SU(3) multiplet structure of QCD in a more realistic way than in the leading-colour approximation, allowing for the formation of “baryonic” configurations where for example two colours can combine coherently to form an anti-colour. The same $\chi^{2}/$ndf approach results in a $78\%$ probability that the model describes the data.
The simulation with PYTHIA~8 with the CR-BLC Mode~2 tune also provides a much more accurate description of the \LcDzero cross section ratio, previously measured in pp collisions at the LHC~\cite{Acharya:2017kfy, Acharya:2020uqi,Acharya:2020lrg,LcD013TeV,ALICE:2021npz,Sirunyan:2019fnc}.

In the right-hand panel of Fig.~\ref{fig:result}, a comparison of the \z distribution of \Lambdac-tagged jets and the \z distribution previously measured for \Dzero-tagged jets~\cite{D0fragmentation} is presented. The latter is consistent with PYTHIA~8 simulations using both the Monash and CR-BLC Mode~2 tunes. The ratio of the two distributions is also presented in the bottom panel. The uncertainty from the jet energy resolution was considered to be correlated between the \Lambdac-tagged jet and \Dzero-tagged jet measurements and was evaluated directly on the ratio of the distributions. The remaining uncertainties were considered uncorrelated when taking the ratio and were then combined in quadrature with the uncertainty of the jet energy resolution. The uncertainties were considered uncorrelated across the \z intervals. The same $\chi^{2}/$ndf exercise described above determines that there is a $12\%$ probability that the measured ratio is described by a flat distribution at unity, hinting at a softer fragmentation of charm quarks into charm baryons than charm mesons. The ratio is better described by the PYTHIA~8 simulations with the CR-BLC Mode~2 compared to the ones with the Monash tune, with the former describing the data with $88\%$ probability compared to a $0.03\%$ probability for the latter.

%% file: 5_Conclusion.tex
In summary the first measurement in hadronic collisions of the longitudinal momentum fraction of the charged jet carried by \Lambdac baryons was presented for \pp collisions at $\s=13$ TeV. The result is fully corrected to particle level and obtained in the jet and \Lambdac transverse\hyp{}momentum intervals of $7 \leq \ptjet < 15$ \GeVc and $3 \leq \ptLc < 15$ \GeVc, respectively. The measurement presented in this paper hints that charm quarks have a softer fragmentation into \Lambdac baryons compared to \Dzero mesons, in the measured kinematic interval. One possible explanation is that charm-baryon production is favoured in the presence of higher particle multiplicity originating from both the jet fragmentation and the underlying event, which could be tested with future measurements of the in-jet multiplicity of \Lambdac-tagged jets. The fragmentation of charm quarks into \Lambdac baryons in hadronic collisions exhibits tension with simulations tuned on \ee data that employ a leading-colour formalism of hadronisation, such as in the Monash tune of PYTHIA~8. This occurs despite their successful description of the fragmentation of charm quarks into \Dzero mesons. However, the inclusion of mechanisms sensitive to the surrounding partonic density that feature colour reconnection beyond the leading-colour approximation results in a better agreement with data. This result also partially explains the $\pt$ shape of the prompt \LcDzero cross section ratio~\cite{Acharya:2017kfy, Acharya:2020uqi,Acharya:2020lrg,LcD013TeV,ALICE:2021npz,Sirunyan:2019fnc}, which shows a peak at low $\pt$ ($\approx 3$~\GeVc) and is also described within uncertainties by PYTHIA 8 with the CR-BLC Mode 2 tune. The $\pt$ trend of this ratio is driven by the fact that the \Lambdac baryons produced from the fragmenting charm quark carry a significantly lower fraction of the charm-quark transverse momentum than the \Dzero mesons produced in a similar way.

%% file: fa_2022-11-21_Opt_C.tex

The ALICE Collaboration would like to thank all its engineers and technicians for their invaluable contributions to the construction of the experiment and the CERN accelerator teams for the outstanding performance of the LHC complex.
The ALICE Collaboration gratefully acknowledges the resources and support provided by all Grid centres and the Worldwide LHC Computing Grid (WLCG) collaboration.
The ALICE Collaboration acknowledges the following funding agencies for their support in building and running the ALICE detector:
A. I. Alikhanyan National Science Laboratory (Yerevan Physics Institute) Foundation (ANSL), State Committee of Science and World Federation of Scientists (WFS), Armenia;
Austrian Academy of Sciences, Austrian Science Fund (FWF): [M 2467-N36] and Nationalstiftung f\"{u}r Forschung, Technologie und Entwicklung, Austria;
Ministry of Communications and High Technologies, National Nuclear Research Center, Azerbaijan;
Conselho Nacional de Desenvolvimento Cient\'{\i}fico e Tecnol\'{o}gico (CNPq), Financiadora de Estudos e Projetos (Finep), Funda\c{c}\~{a}o de Amparo \`{a} Pesquisa do Estado de S\~{a}o Paulo (FAPESP) and Universidade Federal do Rio Grande do Sul (UFRGS), Brazil;
Bulgarian Ministry of Education and Science, within the National Roadmap for Research Infrastructures 2020-2027 (object CERN), Bulgaria;
Ministry of Education of China (MOEC) , Ministry of Science \& Technology of China (MSTC) and National Natural Science Foundation of China (NSFC), China;
Ministry of Science and Education and Croatian Science Foundation, Croatia;
Centro de Aplicaciones Tecnol\'{o}gicas y Desarrollo Nuclear (CEADEN), Cubaenerg\'{\i}a, Cuba;
Ministry of Education, Youth and Sports of the Czech Republic, Czech Republic;
The Danish Council for Independent Research | Natural Sciences, the VILLUM FONDEN and Danish National Research Foundation (DNRF), Denmark;
Helsinki Institute of Physics (HIP), Finland;
Commissariat \`{a} l'Energie Atomique (CEA) and Institut National de Physique Nucl\'{e}aire et de Physique des Particules (IN2P3) and Centre National de la Recherche Scientifique (CNRS), France;
Bundesministerium f\"{u}r Bildung und Forschung (BMBF) and GSI Helmholtzzentrum f\"{u}r Schwerionenforschung GmbH, Germany;
General Secretariat for Research and Technology, Ministry of Education, Research and Religions, Greece;
National Research, Development and Innovation Office, Hungary;
Department of Atomic Energy Government of India (DAE), Department of Science and Technology, Government of India (DST), University Grants Commission, Government of India (UGC) and Council of Scientific and Industrial Research (CSIR), India;
National Research and Innovation Agency - BRIN, Indonesia;
Istituto Nazionale di Fisica Nucleare (INFN), Italy;
Japanese Ministry of Education, Culture, Sports, Science and Technology (MEXT) and Japan Society for the Promotion of Science (JSPS) KAKENHI, Japan;
Consejo Nacional de Ciencia (CONACYT) y Tecnolog\'{i}a, through Fondo de Cooperaci\'{o}n Internacional en Ciencia y Tecnolog\'{i}a (FONCICYT) and Direcci\'{o}n General de Asuntos del Personal Academico (DGAPA), Mexico;
Nederlandse Organisatie voor Wetenschappelijk Onderzoek (NWO), Netherlands;
The Research Council of Norway, Norway;
Commission on Science and Technology for Sustainable Development in the South (COMSATS), Pakistan;
Pontificia Universidad Cat\'{o}lica del Per\'{u}, Peru;
Ministry of Education and Science, National Science Centre and WUT ID-UB, Poland;
Korea Institute of Science and Technology Information and National Research Foundation of Korea (NRF), Republic of Korea;
Ministry of Education and Scientific Research, Institute of Atomic Physics, Ministry of Research and Innovation and Institute of Atomic Physics and Universitatea Nationala de Stiinta si Tehnologie Politehnica Bucuresti, Romania;
Ministry of Education, Science, Research and Sport of the Slovak Republic, Slovakia;
National Research Foundation of South Africa, South Africa;
Swedish Research Council (VR) and Knut \& Alice Wallenberg Foundation (KAW), Sweden;
European Organization for Nuclear Research, Switzerland;
Suranaree University of Technology (SUT), National Science and Technology Development Agency (NSTDA) and National Science, Research and Innovation Fund (NSRF via PMU-B B05F650021), Thailand;
Turkish Energy, Nuclear and Mineral Research Agency (TENMAK), Turkey;
National Academy of  Sciences of Ukraine, Ukraine;
Science and Technology Facilities Council (STFC), United Kingdom;
National Science Foundation of the United States of America (NSF) and United States Department of Energy, Office of Nuclear Physics (DOE NP), United States of America.
In addition, individual groups or members have received support from:
European Research Council, Strong 2020 - Horizon 2020, Marie Sk\l{}odowska Curie (grant nos. 950692, 824093, 896850), European Union;
Academy of Finland (Center of Excellence in Quark Matter) (grant nos. 346327, 346328), Finland;
Programa de Apoyos para la Superaci\'{o}n del Personal Acad\'{e}mico, UNAM, Mexico.

%% file: 2022-11-21-Alice_Authorlist_2022-11-21_Opt_C.tex
\begin{flushleft} 
\small

S.~Acharya\,\orcidlink{0000-0002-9213-5329}\,$^{\rm 126}$, 
D.~Adamov\'{a}\,\orcidlink{0000-0002-0504-7428}\,$^{\rm 86}$, 
A.~Adler$^{\rm 70}$, 
G.~Aglieri Rinella\,\orcidlink{0000-0002-9611-3696}\,$^{\rm 32}$, 
M.~Agnello\,\orcidlink{0000-0002-0760-5075}\,$^{\rm 29}$, 
N.~Agrawal\,\orcidlink{0000-0003-0348-9836}\,$^{\rm 51}$, 
Z.~Ahammed\,\orcidlink{0000-0001-5241-7412}\,$^{\rm 134}$, 
S.~Ahmad\,\orcidlink{0000-0003-0497-5705}\,$^{\rm 15}$, 
S.U.~Ahn\,\orcidlink{0000-0001-8847-489X}\,$^{\rm 71}$, 
I.~Ahuja\,\orcidlink{0000-0002-4417-1392}\,$^{\rm 37}$
% ...

\section*{Affiliation Notes}

$^{\rm I}$ Deceased\\
$^{\rm II}$ Also at: Max-Planck-Institut fur Physik, Munich, Germany\\
$^{\rm III}$ Also at: Italian National Agency for New Technologies, Energy and Sustainable Economic Development (ENEA), Bologna, Italy\\
$^{\rm IV}$ Also at: Dipartimento DET del Politecnico di Torino, Turin, Italy\\
$^{\rm V}$ Also at: Department of Applied Physics, Aligarh Muslim University, Aligarh, India\\
$^{\rm VI}$ Also at: Institute of Theoretical Physics, University of Wroclaw, Poland\\
$^{\rm VII}$ Also at: An institution covered by a cooperation agreement with CERN\\

\section*{Collaboration Institutes}

$^{1}$ A.I. Alikhanyan National Science Laboratory (Yerevan Physics Institute) Foundation, Yerevan, Armenia\\
$^{2}$ AGH University of Krakow, Cracow, Poland\\
$^{3}$ Bogolyubov Institute for Theoretical Physics, National Academy of Sciences of Ukraine, Kiev, Ukraine\\
$^{4}$ Bose Institute, Department of Physics  and Centre for Astroparticle Physics and Space Science (CAPSS), Kolkata, India\\
$^{5}$ California Polytechnic State University, San Luis Obispo, California, United States\\
$^{6}$ Central China Normal University, Wuhan, China\\
$^{7}$ Centro de Aplicaciones Tecnol\'{o}gicas y Desarrollo Nuclear (CEADEN), Havana, Cuba\\
$^{8}$ Centro de Investigaci\'{o}n y de Estudios Avanzados (CINVESTAV), Mexico City and M\'{e}rida, Mexico\\
$^{9}$ Chicago State University, Chicago, Illinois, United States\\
$^{10}$ China Institute of Atomic Energy, Beijing, China\\
$^{11}$ Chungbuk National University, Cheongju, Republic of Korea\\
$^{12}$ Comenius University Bratislava, Faculty of Mathematics, Physics and Informatics, Bratislava, Slovak Republic\\
$^{13}$ COMSATS University Islamabad, Islamabad, Pakistan\\
$^{14}$ Creighton University, Omaha, Nebraska, United States\\
$^{15}$ Department of Physics, Aligarh Muslim University, Aligarh, India\\
$^{16}$ Department of Physics, Pusan National University, Pusan, Republic of Korea\\
$^{17}$ Department of Physics, Sejong University, Seoul, Republic of Korea\\
$^{18}$ Department of Physics, University of California, Berkeley, California, United States\\
$^{19}$ Department of Physics, University of Oslo, Oslo, Norway\\
$^{20}$ Department of Physics and Technology, University of Bergen, Bergen, Norway\\
$^{21}$ Dipartimento di Fisica, Universit\`{a} di Pavia, Pavia, Italy\\
$^{22}$ Dipartimento di Fisica dell'Universit\`{a} and Sezione INFN, Cagliari, Italy\\
$^{23}$ Dipartimento di Fisica dell'Universit\`{a} and Sezione INFN, Trieste, Italy\\
$^{24}$ Dipartimento di Fisica dell'Universit\`{a} and Sezione INFN, Turin, Italy\\
$^{25}$ Dipartimento di Fisica e Astronomia dell'Universit\`{a} and Sezione INFN, Bologna, Italy\\
$^{26}$ Dipartimento di Fisica e Astronomia dell'Universit\`{a} and Sezione INFN, Catania, Italy\\
$^{27}$ Dipartimento di Fisica e Astronomia dell'Universit\`{a} and Sezione INFN, Padova, Italy\\
$^{28}$ Dipartimento di Fisica `E.R.~Caianiello' dell'Universit\`{a} and Gruppo Collegato INFN, Salerno, Italy\\
$^{29}$ Dipartimento DISAT del Politecnico and Sezione INFN, Turin, Italy\\
$^{30}$ Dipartimento di Scienze MIFT, Universit\`{a} di Messina, Messina, Italy\\
$^{31}$ Dipartimento Interateneo di Fisica `M.~Merlin' and Sezione INFN, Bari, Italy\\
$^{32}$ European Organization for Nuclear Research (CERN), Geneva, Switzerland\\
$^{33}$ Faculty of Electrical Engineering, Mechanical Engineering and Naval Architecture, University of Split, Split, Croatia\\
$^{34}$ Faculty of Engineering and Science, Western Norway University of Applied Sciences, Bergen, Norway\\
$^{35}$ Faculty of Nuclear Sciences and Physical Engineering, Czech Technical University in Prague, Prague, Czech Republic\\
$^{36}$ Faculty of Physics, Sofia University, Sofia, Bulgaria\\
$^{37}$ Faculty of Science, P.J.~\v{S}af\'{a}rik University, Ko\v{s}ice, Slovak Republic\\
$^{38}$ Frankfurt Institute for Advanced Studies, Johann Wolfgang Goethe-Universit\"{a}t Frankfurt, Frankfurt, Germany\\
$^{39}$ Fudan University, Shanghai, China\\
$^{40}$ Gangneung-Wonju National University, Gangneung, Republic of Korea\\
$^{41}$ Gauhati University, Department of Physics, Guwahati, India\\
$^{42}$ Helmholtz-Institut f\"{u}r Strahlen- und Kernphysik, Rheinische Friedrich-Wilhelms-Universit\"{a}t Bonn, Bonn, Germany\\
$^{43}$ Helsinki Institute of Physics (HIP), Helsinki, Finland\\
$^{44}$ High Energy Physics Group,  Universidad Aut\'{o}noma de Puebla, Puebla, Mexico\\
$^{45}$ Horia Hulubei National Institute of Physics and Nuclear Engineering, Bucharest, Romania\\
$^{46}$ HUN-REN Wigner Research Centre for Physics, Budapest, Hungary\\
$^{47}$ Indian Institute of Technology Bombay (IIT), Mumbai, India\\
$^{48}$ Indian Institute of Technology Indore, Indore, India\\
$^{49}$ INFN, Laboratori Nazionali di Frascati, Frascati, Italy\\
$^{50}$ INFN, Sezione di Bari, Bari, Italy\\
$^{51}$ INFN, Sezione di Bologna, Bologna, Italy\\
$^{52}$ INFN, Sezione di Cagliari, Cagliari, Italy\\
$^{53}$ INFN, Sezione di Catania, Catania, Italy\\
$^{54}$ INFN, Sezione di Padova, Padova, Italy\\
$^{55}$ INFN, Sezione di Pavia, Pavia, Italy\\
$^{56}$ INFN, Sezione di Torino, Turin, Italy\\
$^{57}$ INFN, Sezione di Trieste, Trieste, Italy\\
$^{58}$ Inha University, Incheon, Republic of Korea\\
$^{59}$ Institute for Gravitational and Subatomic Physics (GRASP), Utrecht University/Nikhef, Utrecht, Netherlands\\
$^{60}$ Institute of Experimental Physics, Slovak Academy of Sciences, Ko\v{s}ice, Slovak Republic\\
$^{61}$ Institute of Physics, Homi Bhabha National Institute, Bhubaneswar, India\\
$^{62}$ Institute of Physics of the Czech Academy of Sciences, Prague, Czech Republic\\
$^{63}$ Institute of Space Science (ISS), Bucharest, Romania\\
$^{64}$ Institut f\"{u}r Kernphysik, Johann Wolfgang Goethe-Universit\"{a}t Frankfurt, Frankfurt, Germany\\
$^{65}$ Instituto de Ciencias Nucleares, Universidad Nacional Aut\'{o}noma de M\'{e}xico, Mexico City, Mexico\\
$^{66}$ Instituto de F\'{i}sica, Universidade Federal do Rio Grande do Sul (UFRGS), Porto Alegre, Brazil\\
$^{67}$ Instituto de F\'{\i}sica, Universidad Nacional Aut\'{o}noma de M\'{e}xico, Mexico City, Mexico\\
$^{68}$ iThemba LABS, National Research Foundation, Somerset West, South Africa\\
$^{69}$ Jeonbuk National University, Jeonju, Republic of Korea\\
$^{70}$ Johann-Wolfgang-Goethe Universit\"{a}t Frankfurt Institut f\"{u}r Informatik, Fachbereich Informatik und Mathematik, Frankfurt, Germany\\
$^{71}$ Korea Institute of Science and Technology Information, Daejeon, Republic of Korea\\
$^{72}$ KTO Karatay University, Konya, Turkey\\
$^{73}$ Laboratoire de Physique Subatomique et de Cosmologie, Universit\'{e} Grenoble-Alpes, CNRS-IN2P3, Grenoble, France\\
$^{74}$ Lawrence Berkeley National Laboratory, Berkeley, California, United States\\
$^{75}$ Lund University Department of Physics, Division of Particle Physics, Lund, Sweden\\
$^{76}$ Nagasaki Institute of Applied Science, Nagasaki, Japan\\
$^{77}$ Nara Women{'}s University (NWU), Nara, Japan\\
$^{78}$ National and Kapodistrian University of Athens, School of Science, Department of Physics , Athens, Greece\\
$^{79}$ National Centre for Nuclear Research, Warsaw, Poland\\
$^{80}$ National Institute of Science Education and Research, Homi Bhabha National Institute, Jatni, India\\
$^{81}$ National Nuclear Research Center, Baku, Azerbaijan\\
$^{82}$ National Research and Innovation Agency - BRIN, Jakarta, Indonesia\\
$^{83}$ Niels Bohr Institute, University of Copenhagen, Copenhagen, Denmark\\
$^{84}$ Nikhef, National institute for subatomic physics, Amsterdam, Netherlands\\
$^{85}$ Nuclear Physics Group, STFC Daresbury Laboratory, Daresbury, United Kingdom\\
$^{86}$ Nuclear Physics Institute of the Czech Academy of Sciences, Husinec-\v{R}e\v{z}, Czech Republic\\
$^{87}$ Oak Ridge National Laboratory, Oak Ridge, Tennessee, United States\\
$^{88}$ Ohio State University, Columbus, Ohio, United States\\
$^{89}$ Physics department, Faculty of science, University of Zagreb, Zagreb, Croatia\\
$^{90}$ Physics Department, Panjab University, Chandigarh, India\\
$^{91}$ Physics Department, University of Jammu, Jammu, India\\
$^{92}$ Physics Program and International Institute for Sustainability with Knotted Chiral Meta Matter (SKCM2), Hiroshima University, Hiroshima, Japan\\
$^{93}$ Physikalisches Institut, Eberhard-Karls-Universit\"{a}t T\"{u}bingen, T\"{u}bingen, Germany\\
$^{94}$ Physikalisches Institut, Ruprecht-Karls-Universit\"{a}t Heidelberg, Heidelberg, Germany\\
$^{95}$ Physik Department, Technische Universit\"{a}t M\"{u}nchen, Munich, Germany\\
$^{96}$ Politecnico di Bari and Sezione INFN, Bari, Italy\\
$^{97}$ Research Division and ExtreMe Matter Institute EMMI, GSI Helmholtzzentrum f\"ur Schwerionenforschung GmbH, Darmstadt, Germany\\
$^{98}$ Saga University, Saga, Japan\\
$^{99}$ Saha Institute of Nuclear Physics, Homi Bhabha National Institute, Kolkata, India\\
$^{100}$ School of Physics and Astronomy, University of Birmingham, Birmingham, United Kingdom\\
$^{101}$ Secci\'{o}n F\'{\i}sica, Departamento de Ciencias, Pontificia Universidad Cat\'{o}lica del Per\'{u}, Lima, Peru\\
$^{102}$ Stefan Meyer Institut f\"{u}r Subatomare Physik (SMI), Vienna, Austria\\
$^{103}$ SUBATECH, IMT Atlantique, Nantes Universit\'{e}, CNRS-IN2P3, Nantes, France\\
$^{104}$ Sungkyunkwan University, Suwon City, Republic of Korea\\
$^{105}$ Suranaree University of Technology, Nakhon Ratchasima, Thailand\\
$^{106}$ Technical University of Ko\v{s}ice, Ko\v{s}ice, Slovak Republic\\
$^{107}$ The Henryk Niewodniczanski Institute of Nuclear Physics, Polish Academy of Sciences, Cracow, Poland\\
$^{108}$ The University of Texas at Austin, Austin, Texas, United States\\
$^{109}$ Universidad Aut\'{o}noma de Sinaloa, Culiac\'{a}n, Mexico\\
$^{110}$ Universidade de S\~{a}o Paulo (USP), S\~{a}o Paulo, Brazil\\
$^{111}$ Universidade Estadual de Campinas (UNICAMP), Campinas, Brazil\\
$^{112}$ Universidade Federal do ABC, Santo Andre, Brazil\\
$^{113}$ Universitatea Nationala de Stiinta si Tehnologie Politehnica Bucuresti, Bucharest, Romania\\
$^{114}$ University of Cape Town, Cape Town, South Africa\\
$^{115}$ University of Houston, Houston, Texas, United States\\
$^{116}$ University of Jyv\"{a}skyl\"{a}, Jyv\"{a}skyl\"{a}, Finland\\
$^{117}$ University of Kansas, Lawrence, Kansas, United States\\
$^{118}$ University of Liverpool, Liverpool, United Kingdom\\
$^{119}$ University of Science and Technology of China, Hefei, China\\
$^{120}$ University of South-Eastern Norway, Kongsberg, Norway\\
$^{121}$ University of Tennessee, Knoxville, Tennessee, United States\\
$^{122}$ University of the Witwatersrand, Johannesburg, South Africa\\
$^{123}$ University of Tokyo, Tokyo, Japan\\
$^{124}$ University of Tsukuba, Tsukuba, Japan\\
$^{125}$ Universit\"{a}t M\"{u}nster, Institut f\"{u}r Kernphysik, M\"{u}nster, Germany\\
$^{126}$ Universit\'{e} Clermont Auvergne, CNRS/IN2P3, LPC, Clermont-Ferrand, France\\
$^{127}$ Universit\'{e} de Lyon, CNRS/IN2P3, Institut de Physique des 2 Infinis de Lyon, Lyon, France\\
$^{128}$ Universit\'{e} de Strasbourg, CNRS, IPHC UMR 7178, F-67000 Strasbourg, France, Strasbourg, France\\
$^{129}$ Universit\'{e} Paris-Saclay, Centre d'Etudes de Saclay (CEA), IRFU, D\'{e}partment de Physique Nucl\'{e}aire (DPhN), Saclay, France\\
$^{130}$ Universit\'{e}  Paris-Saclay, CNRS/IN2P3, IJCLab, Orsay, France\\
$^{131}$ Universit\`{a} degli Studi di Foggia, Foggia, Italy\\
$^{132}$ Universit\`{a} del Piemonte Orientale, Vercelli, Italy\\
$^{133}$ Universit\`{a} di Brescia, Brescia, Italy\\
$^{134}$ Variable Energy Cyclotron Centre, Homi Bhabha National Institute, Kolkata, India\\
$^{135}$ Warsaw University of Technology, Warsaw, Poland\\
$^{136}$ Wayne State University, Detroit, Michigan, United States\\
$^{137}$ Yale University, New Haven, Connecticut, United States\\
$^{138}$ Yonsei University, Seoul, Republic of Korea\\
$^{139}$  Zentrum  f\"{u}r Technologie und Transfer (ZTT), Worms, Germany\\
$^{140}$ Affiliated with an institute covered by a cooperation agreement with CERN\\
$^{141}$ Affiliated with an international laboratory covered by a cooperation agreement with CERN.\\

\end{flushleft} 

%% file: main.bbl
\providecommand{\href}[2]{#2}\begingroup\raggedright\begin{thebibliography}{10}

\bibitem{Collins:1989gx}
J.~C. Collins, D.~E. Soper, and G.~F. Sterman, ``{Factorization of Hard
  Processes in QCD}'', \href{http://dx.doi.org/10.1142/9789814503266_0001}{{\em
  Adv. Ser. Direct. High Energy Phys.} {\bfseries 5} (1989) 1--91},
  \href{http://arxiv.org/abs/hep-ph/0409313}{{\ttfamily arXiv:hep-ph/0409313}}.

\bibitem{belle_charm_fragmentation}
{\bfseries Belle} Collaboration, R.~Seuster {\em et~al.}, ``{Charm hadrons from
  fragmentation and B decays in $\textnormal{e}^{+}\textnormal{e}^{-}$
  annihilation at $\sqrt{s}=10.6$~\GeV}'',
  \href{http://dx.doi.org/10.1103/PhysRevD.73.032002}{{\em Phys. Rev. D}
  {\bfseries 73} (2006) 032002},
  \href{http://arxiv.org/abs/hep-ex/0506068}{{\ttfamily arXiv:hep-ex/0506068}}.

\bibitem{belle_charm_crosssection}
{\bfseries Belle} Collaboration, M.~Niiyama {\em et~al.}, ``{Production cross
  sections of hyperons and charmed baryons from
  $\textnormal{e}^{+}\textnormal{e}^{-}$ annihilation near $\sqrt{s} =
  10.52$~\GeV}'', \href{http://dx.doi.org/10.1103/PhysRevD.97.072005}{{\em
  Phys. Rev. D} {\bfseries 97} (2018) 072005},
  \href{http://arxiv.org/abs/1706.06791}{{\ttfamily arXiv:1706.06791
  [hep-ex]}}.

\bibitem{aleph_charm_production}
{\bfseries ALEPH} Collaboration, R.~Barate {\em et~al.}, ``{Study of charm
  production in Z decays}'',
  \href{http://dx.doi.org/10.1007/s100520000421}{{\em Eur. Phys. J. C}
  {\bfseries 16} (2000) 597--611},
  \href{http://arxiv.org/abs/hep-ex/9909032}{{\ttfamily arXiv:hep-ex/9909032}}.

\bibitem{zeus_charm_fragmentation}
{\bfseries ZEUS} Collaboration, S.~Chekanov {\em et~al.}, ``{Measurement of the
  charm fragmentation function in D* photoproduction at HERA}'',
  \href{http://dx.doi.org/10.1088/1126-6708/2009/04/082}{{\em JHEP} {\bfseries
  04} (2009) 082}, \href{http://arxiv.org/abs/0901.1210}{{\ttfamily
  arXiv:0901.1210 [hep-ex]}}.

\bibitem{H1:2008nbh}
{\bfseries H1} Collaboration, F.~D. Aaron {\em et~al.}, ``{Study of Charm
  Fragmentation into D*+- Mesons in Deep-Inelastic Scattering at HERA}'',
  \href{http://dx.doi.org/10.1140/epjc/s10052-008-0792-2}{{\em Eur. Phys. J. C}
  {\bfseries 59} (2009) 589--606},
  \href{http://arxiv.org/abs/0808.1003}{{\ttfamily arXiv:0808.1003 [hep-ex]}}.

\bibitem{D0fragmentation}
{\bfseries ALICE} Collaboration, S.~Acharya {\em et~al.}, ``{Measurement of the
  production of charm jets tagged with $\Dzero$ mesons in pp collisions at
  $\sqrt{s} = 5.02$ and $13~\TeV$}'',
  \href{http://dx.doi.org/10.1007/JHEP06(2023)133}{{\em JHEP} {\bfseries 06}
  (2023) 133}, \href{http://arxiv.org/abs/2204.10167}{{\ttfamily
  arXiv:2204.10167 [nucl-ex]}}.

\bibitem{D0fragmentation7Tev}
{\bfseries ALICE} Collaboration, S.~Acharya {\em et~al.}, ``{Measurement of the
  production of charm jets tagged with $\Dzero$ mesons in pp collisions at
  $\sqrt{s}=7~\TeV$}'', \href{http://dx.doi.org/10.1007/JHEP08(2019)133}{{\em
  JHEP} {\bfseries 08} (2019) 133},
  \href{http://arxiv.org/abs/1905.02510}{{\ttfamily arXiv:1905.02510
  [nucl-ex]}}.

\bibitem{ATLAS:2011chi}
{\bfseries ATLAS} Collaboration, G.~Aad {\em et~al.}, ``{Measurement of
  D$^{*\pm}$ meson production in jets from pp collisions at $\sqrt{s} = 7~\TeV$
  with the ATLAS detector}'',
  \href{http://dx.doi.org/10.1103/PhysRevD.85.052005}{{\em Phys. Rev. D}
  {\bfseries 85} (2012) 052005},
  \href{http://arxiv.org/abs/1112.4432}{{\ttfamily arXiv:1112.4432 [hep-ex]}}.

\bibitem{ALICE:2016fzo}
{\bfseries ALICE} Collaboration, J.~Adam {\em et~al.}, ``{Enhanced production
  of multi-strange hadrons in high-multiplicity proton-proton collisions}'',
  \href{http://dx.doi.org/10.1038/nphys4111}{{\em Nature Phys.} {\bfseries 13}
  (2017) 535--539}, \href{http://arxiv.org/abs/1606.07424}{{\ttfamily
  arXiv:1606.07424 [nucl-ex]}}.

\bibitem{Acharya:2017kfy}
{\bfseries ALICE} Collaboration, S.~Acharya {\em et~al.}, ``{\Lc production in
  pp collisions at $\sqrt{s} = 7$ TeV and in p--Pb collisions at $\sqrt{s_{\rm
  NN}} = 5.02$ TeV}'', \href{http://dx.doi.org/10.1007/JHEP04(2018)108}{{\em
  JHEP} {\bfseries 04} (2018) 108},
\href{http://arxiv.org/abs/1712.09581}{{\ttfamily arXiv:1712.09581 [nucl-ex]}}.

\bibitem{Acharya:2020uqi}
{\bfseries ALICE} Collaboration, S.~Acharya {\em et~al.}, ``{\Lc Production and
  Baryon-to-Meson Ratios in pp and p--Pb Collisions at $\sqrt
  {s_{\textnormal{NN}}} = 5.02$ TeV at the LHC}'',
  \href{http://dx.doi.org/10.1103/PhysRevLett.127.202301}{{\em Phys. Rev.
  Lett.} {\bfseries 127} (2021) 202301},
  \href{http://arxiv.org/abs/2011.06078}{{\ttfamily arXiv:2011.06078
  [nucl-ex]}}.

\bibitem{LcD013TeV}
{\bfseries ALICE} Collaboration, S.~Acharya {\em et~al.}, ``{Measurement of
  Prompt D$^{0}$, \Lc, and $\upSigma_{\textnormal{c}}^{0,++}$(2455) Production
  in Proton--Proton Collisions at $\sqrt{s} = 13$ TeV}'',
  \href{http://dx.doi.org/10.1103/PhysRevLett.128.012001}{{\em Phys. Rev.
  Lett.} {\bfseries 128} (2022) 012001},
  \href{http://arxiv.org/abs/2106.08278}{{\ttfamily arXiv:2106.08278
  [hep-ex]}}.

\bibitem{Acharya:2020lrg}
{\bfseries ALICE} Collaboration, S.~Acharya {\em et~al.}, ``{\Lc production in
  pp and in p--Pb collisions at $\sqrt {s_{\textnormal{NN}}} = 5.02$ TeV}'',
  \href{http://dx.doi.org/10.1103/PhysRevC.104.054905}{{\em Phys. Rev. C}
  {\bfseries 104} (2021) 054905},
  \href{http://arxiv.org/abs/2011.06079}{{\ttfamily arXiv:2011.06079
  [nucl-ex]}}.

\bibitem{ALICE:2021npz}
{\bfseries ALICE} Collaboration, S.~Acharya {\em et~al.}, ``{Observation of a
  multiplicity dependence in the $\pt$-differential charm baryon-to-meson
  ratios in proton--proton collisions at $\sqrt{s} = 13$ TeV}'',
  \href{http://dx.doi.org/10.1016/j.physletb.2022.137065}{{\em Phys. Lett. B}
  {\bfseries 829} (2022) 137065},
  \href{http://arxiv.org/abs/2111.11948}{{\ttfamily arXiv:2111.11948
  [nucl-ex]}}.

\bibitem{ALICE:2021dhb}
{\bfseries ALICE} Collaboration, S.~Acharya {\em et~al.}, ``{Charm-quark
  fragmentation fractions and production cross section at midrapidity in pp
  collisions at the LHC}'',
  \href{http://dx.doi.org/10.1103/PhysRevD.105.L011103}{{\em Phys. Rev. D}
  {\bfseries 105} (2022) L011103},
  \href{http://arxiv.org/abs/2105.06335}{{\ttfamily arXiv:2105.06335
  [nucl-ex]}}.

\bibitem{ALICE:OmegaC}
{\bfseries ALICE} Collaboration, S.~Acharya {\em et~al.}, ``{First measurement
  of $\mathrm{\Omega_{\textnormal{c}}^{0}}$ production in pp collisions at
  $\sqrt{s}=13$ TeV}'',
  \href{http://dx.doi.org/10.1016/j.physletb.2022.137625}{{\em Phys. Lett. B}
  {\bfseries 846} (2023) 137625},
  \href{http://arxiv.org/abs/2205.13993}{{\ttfamily arXiv:2205.13993
  [nucl-ex]}}.

\bibitem{ALICE:XiC}
{\bfseries ALICE} Collaboration, S.~Acharya {\em et~al.}, ``{First measurement
  of $\mathrm{\Xi_{c}^{0}}$ production in pp collisions at $\sqrt{s} = 7$
  TeV}'', \href{http://dx.doi.org/10.1016/j.physletb.2018.03.061}{{\em Phys.
  Lett. B} {\bfseries 781} (2018) 8--19},
  \href{http://arxiv.org/abs/1712.04242}{{\ttfamily arXiv:1712.04242
  [hep-ex]}}.

\bibitem{ALICE:XiC2}
{\bfseries ALICE} Collaboration, S.~Acharya {\em et~al.}, ``{Measurement of the
  Cross Sections of $\Xi^0_{c}$ and $\Xi^+_{c}$ Baryons and of the
  Branching-Fraction Ratio BR($\Xi^0_{c} \rightarrow \Xi^-{e}^+\nu_{
  e}$)/BR($\Xi^0_{c} \rightarrow \Xi^-\pi^+$) in pp collisions at 13 TeV}'',
  \href{http://dx.doi.org/10.1103/PhysRevLett.127.272001}{{\em Phys. Rev.
  Lett.} {\bfseries 127} (2021) 272001},
  \href{http://arxiv.org/abs/2105.05187}{{\ttfamily arXiv:2105.05187
  [nucl-ex]}}.

\bibitem{Skands:2014pea}
P.~Skands, S.~Carrazza, and J.~Rojo, ``{Tuning PYTHIA 8.1: the Monash 2013
  Tune}'', \href{http://dx.doi.org/10.1140/epjc/s10052-014-3024-y}{{\em Eur.
  Phys. J. C} {\bfseries 74} (2014) 3024},
\href{http://arxiv.org/abs/1404.5630}{{\ttfamily arXiv:1404.5630 [hep-ph]}}.

\bibitem{Sjostrand:2014zea}
T.~Sj{\"o}strand {\em et~al.}, ``{An introduction to PYTHIA 8.2}'',
  \href{http://dx.doi.org/10.1016/j.cpc.2015.01.024}{{\em Comput. Phys.
  Commun.} {\bfseries 191} (2015) 159--177},
  \href{http://arxiv.org/abs/1410.3012}{{\ttfamily arXiv:1410.3012 [hep-ph]}}.

\bibitem{Bahr:2008pv}
M.~Bahr {\em et~al.}, ``{Herwig++ Physics and Manual}'',
  \href{http://dx.doi.org/10.1140/epjc/s10052-008-0798-9}{{\em Eur. Phys. J. C}
  {\bfseries 58} (2008) 639--707},
\href{http://arxiv.org/abs/0803.0883}{{\ttfamily arXiv:0803.0883 [hep-ph]}}.

\bibitem{Christiansen:2015yqa}
J.~R. Christiansen and P.~Z. Skands, ``{String Formation Beyond Leading
  Colour}'', \href{http://dx.doi.org/10.1007/JHEP08(2015)003}{{\em JHEP}
  {\bfseries 08} (2015) 003},
\href{http://arxiv.org/abs/1505.01681}{{\ttfamily arXiv:1505.01681 [hep-ph]}}.

\bibitem{Aamodt:2008zz}
{\bfseries ALICE} Collaboration, K.~Aamodt {\em et~al.}, ``{The ALICE
  experiment at the CERN LHC}'',
  \href{http://dx.doi.org/10.1088/1748-0221/3/08/S08002}{{\em JINST} {\bfseries
  3} (2008) S08002}.

\bibitem{Abelev:2014ffa}
{\bfseries ALICE} Collaboration, B.~B. Abelev {\em et~al.}, ``{Performance of
  the ALICE Experiment at the CERN LHC}'',
  \href{http://dx.doi.org/10.1142/S0217751X14300440}{{\em Int. J. Mod. Phys. A}
  {\bfseries 29} (2014) 1430044},
  \href{http://arxiv.org/abs/1402.4476}{{\ttfamily arXiv:1402.4476 [nucl-ex]}}.

\bibitem{alice-luminosity}
{\bfseries ALICE} Collaboration, S.~Acharya {\em et~al.}, ``{ALICE
  2016-2017-2018 luminosity determination for pp collisions at $\sqrt{s} =
  13~\TeV$}'',. \url{https://cds.cern.ch/record/2776672}.

\bibitem{Zyla:2020zbs}
{\bfseries Particle Data Group} Collaboration, P.~Zyla {\em et~al.}, ``{Review
  of Particle Physics}'', \href{http://dx.doi.org/10.1093/ptep/ptaa104}{{\em
  PTEP} {\bfseries 2020} (2020) 083C01}.

\bibitem{Chen:2016:XST:2939672.2939785}
T.~Chen and C.~Guestrin,
  \href{http://dx.doi.org/10.1145/2939672.2939785}{``{XGBoost}: A scalable tree
  boosting system'',} in {\em Proceedings of the 22nd ACM SIGKDD International
  Conference on Knowledge Discovery and Data Mining}, KDD '16, pp.~785--794.
\newblock ACM, 2016.
\newblock \href{http://arxiv.org/abs/1603.02754}{{\ttfamily arXiv:1603.02754
  [cs.LG]}}.

\bibitem{GEANT3}
R.~Brun, F.~Bruyant, M.~Maire, A.~C. McPherson, and P.~Zanarini, {\em {GEANT 3
  : user's guide Geant 3.10, Geant 3.11; rev. version}}.
\newblock CERN, Geneva, 1987.
\newblock \url{https://cds.cern.ch/record/1119728}.

\bibitem{Cacciari_2012}
M.~Cacciari, G.~P. Salam, and G.~Soyez, ``{FastJet user manual}'',
  \href{http://dx.doi.org/10.1140/epjc/s10052-012-1896-2}{{\em Eur. Phys. J. C}
  {\bfseries 72} (2012) 1896}, \href{http://arxiv.org/abs/1111.6097}{{\ttfamily
  arXiv:1111.6097 [hep-ph]}}.

\bibitem{Cacciari_2008}
M.~Cacciari, G.~P. Salam, and G.~Soyez, ``{The anti-$k_{\textnormal{t}}$ jet
  clustering algorithm}'',
  \href{http://dx.doi.org/10.1088/1126-6708/2008/04/063}{{\em JHEP} {\bfseries
  04} (2008) 063}, \href{http://arxiv.org/abs/0802.1189}{{\ttfamily
  arXiv:0802.1189 [hep-ph]}}.

\bibitem{EScheme}
M.~Cacciari, G.~P. Salam, and G.~Soyez, ``{The Catchment Area of Jets}'',
  \href{http://dx.doi.org/10.1088/1126-6708/2008/04/005}{{\em JHEP} {\bfseries
  04} (2008) 005}, \href{http://arxiv.org/abs/0802.1188}{{\ttfamily
  arXiv:0802.1188 [hep-ph]}}.

\bibitem{Han:2016uhh}
K.~C. Han, R.~J. Fries, and C.~M. Ko, ``{Jet Fragmentation via Recombination of
  Parton Showers}'', \href{http://dx.doi.org/10.1103/PhysRevC.93.045207}{{\em
  Phys. Rev. C} {\bfseries 93} (2016) 045207},
  \href{http://arxiv.org/abs/1601.00708}{{\ttfamily arXiv:1601.00708
  [nucl-th]}}.

\bibitem{POWHEG}
S.~Alioli, P.~Nason, C.~Oleari, and E.~Re, ``{A general framework for
  implementing NLO calculations in shower Monte Carlo programs: the POWHEG
  BOX}'', \href{http://dx.doi.org/10.1007/JHEP06(2010)043}{{\em JHEP}
  {\bfseries 06} (2010) 043}, \href{http://arxiv.org/abs/1002.2581}{{\ttfamily
  arXiv:1002.2581 [hep-ph]}}.

\bibitem{Sjostrand:2006za}
T.~Sj{\"o}strand, S.~Mrenna, and P.~Z. Skands, ``{PYTHIA 6.4 Physics and
  Manual}'', \href{http://dx.doi.org/10.1088/1126-6708/2006/05/026}{{\em JHEP}
  {\bfseries 05} (2006) 026},
  \href{http://arxiv.org/abs/hep-ph/0603175}{{\ttfamily arXiv:hep-ph/0603175}}.

\bibitem{evtgen}
D.~J. Lange, ``{The EvtGen particle decay simulation package}'',
  \href{http://dx.doi.org/https://doi.org/10.1016/S0168-9002(01)00089-4}{{\em
  Nuclear Instruments and Methods in Physics Research Section A: Accelerators,
  Spectrometers, Detectors and Associated Equipment} {\bfseries 462} (2001)
  152--155}.

\bibitem{unfolding}
G.~D'Agostini, ``{A multidimensional unfolding method based on Bayes'
  theorem}'', \href{http://dx.doi.org/10.1016/0168-9002(95)00274-X}{{\em Nucl.
  Instrum. Meth. A} {\bfseries 362} (1995) 487--498}.

\bibitem{powheg-variations}
M.~Cacciari, P.~Nason, and R.~Vogt, ``{QCD predictions for charm and bottom
  quark production at RHIC}'',
  \href{http://dx.doi.org/10.1103/PhysRevLett.95.122001}{{\em Phys. Rev. Lett.}
  {\bfseries 95} (2005) 122001},
  \href{http://arxiv.org/abs/hep-ph/0502203}{{\ttfamily arXiv:hep-ph/0502203}}.

\bibitem{Sirunyan:2019fnc}
{\bfseries CMS} Collaboration, A.~M. Sirunyan {\em et~al.}, ``{Production of
  \Lc baryons in proton-proton and lead-lead collisions at
  $\sqrt{s_\mathrm{NN}}= 5.02$ TeV}'',
  \href{http://dx.doi.org/10.1016/j.physletb.2020.135328}{{\em Phys. Lett. B}
  {\bfseries 803} (2020) 135328},
  \href{http://arxiv.org/abs/1906.03322}{{\ttfamily arXiv:1906.03322
  [hep-ex]}}.

\end{thebibliography}\endgroup
